\documentclass[prl, showpacs, twocolumn]{revtex4}
\usepackage{graphics}

\begin{document}

\title{Phenomenological framework for nonequilibrium steady states of molecular motors}
\author{Takahiro Harada}
\email[Electronic address: ]{harada@chem.scphys.kyoto-u.ac.jp}
\affiliation{Department of Physics, Graduate School of Science, Kyoto University \& CREST, Kyoto 606-8502, Japan}

\begin{abstract}
An expression for the energetic efficiency of a molecular motor is presented in terms of an effective temperature, which was defined based on the ratio of the correlation function to the susceptibility of its velocity.
We also present a numerical result regarding this temperature for a thermal ratchet model.
Furthermore, our expression of efficiency using known values for kinesin can adequately explain the experimental results.
The framework presented here has a closed form phenomenologically, and is independent of any detailed models.
\end{abstract}


\maketitle

What is the essential difference between our muscle and the Carnot cycle as an energy transducer?
The recent development of single-molecule-detection techniques \cite{Block} for molecular motors led to new perspectives for this question, by directly visualizing their noisy way of chemo-mechanical energy transduction at the single-molecule level.
Many theoretical studies have been made based on such experiments to present various models \cite{Reimann}, upon which energetics have also been discussed \cite{Parrondo}.
However, since most of them are bottom-up type theories, they are in a dilemma that they cannot be compared with experiments until a model and parameters for a molecular motor have been determined completely.
This is the very difference from a top-down type theory, namely a phenomenology such as thermodynamics, which deals with energetics based on macroscopic observables alone.
It thus sounds meaningful to construct a phenomenology for a molecular motor.

In this Letter, we present an expression for the energy efficiency of a molecular motor based on the susceptibility and the correlation function of its velocity, both of which are observables.
For this purpose, we introduce an effective temperature as the ratio of the correlation function to the susceptibility.
According to Fluctuation Dissipation Theorem (FDT) \cite{Kubo}, this effective temperature coincides the temperature of the heat bath when the motor is in equilibrium, while we show that it varies in general with respect to the timescale of measurement, if the motor is at a nonequilibrium steady state.

First, we define the effective temperature.
The velocity $v(t)$ of a colloidal particle in solution responds linearly to a small external force $f(t)$ as
\begin{equation}
\langle v(t) \rangle - v_0 = \int_{-\infty}^{t} \chi(t - t') f(t') \mathrm{d}t',
\label{eq:response}
\end{equation}
where $\langle ... \rangle$ is the statistical average, $v_0$ is the mean velocity of a reference system (a steady system without perturbation), and $\chi(t)$ is called a response function.
The Fourier-Laplace transformation of $\chi(t)$, given as
\begin{equation}
\chi [\omega] = \int_{0}^{\infty} \chi(t) e^{i \omega t} \mathrm{d}t,
\label{eq:susceptibility}
\end{equation}
is called the susceptibility.
The 1st type of FDT states that if the reference system is in equilibrium, the real part of the susceptibility $\chi' [\omega] \equiv \mathrm{Re} \chi [\omega]$ is proportional to the correlation function of the fluctuation of the reference system.
\begin{equation}
\Phi [\omega] = T_0 \chi' [\omega],
\label{eq:FDT}
\end{equation}
where
\begin{equation}
\Phi [\omega] = \frac{1}{2} \int_{-\infty}^{\infty} \langle (v(0) - v_0) (v(t) - v_0) \rangle e^{i \omega t} \mathrm{d} t.
\label{eq:correlation}
\end{equation}
$T_0$ is the temperature of the reference equilibrium system. We set the Boltzmann constant to unity.

We proceed to the case where the reference system is not in equilibrium but rather in a nonequilibrium steady state.
In this case, since FDT may not hold in a normal sense, the coefficient in Eq. (\ref{eq:FDT}) differs from usual temperature in general even when the system is in contact with a single heat bath.
We can adopt Eq. (\ref{eq:FDT}) as a definition of temperature for the nonequilibrium steady state.
After calculating the susceptibility (Eq. (\ref{eq:susceptibility})) and the correlation function (Eq. (\ref{eq:correlation})) for a nonequilibrium steady state as a reference system, the effective temperature is obtained as
\begin{equation}
T [\omega] \equiv \frac{\Phi [\omega]}{\chi' [\omega]}.
\label{eq:FDT_temp}
\end{equation}
We call this a `dissipation temperature'.
This formulation for a degree of freedom asymmetric with respect to a time reversal, such as velocity, is a counterpart of that introduced in spin glass systems \cite{Peliti} for a symmetrical degree of freedom such as magnetization.
It may be obvious that the dissipation temperature varies depending on the timescale $\omega$ of observation in general.

Before an energetics argument, we compute the dissipation temperature for a typical model of thermal ratchet, known as a flashing ratchet \cite{Astumian, Prost}, in a nonequilibrium steady state, to gain insight into this newly defined temperature.
A flashing ratchet is one of the well known models for molecular motors \cite{Okada}, in which Brownian motion of a particle can be rectified in one direction at a nonequilibrium steady state.
In this paper, we suppose that a particle has two chemical states, each of which provides a periodic potential profile $U_1 (x)$ and $U_2(x)$, respectively (see the inset in Fig. 1 (a)).
The transition between these two states is assumed to be stochastic at the same and constant transition rate $\alpha$.
The Langevin equation and the transition dynamics for the particle are
\begin{equation}
\begin{array}{c}
m \ddot{x} + \gamma \dot{x} = - \partial_x U_i(x) + f(t) + \xi (t), \\
\langle \xi (t) \rangle = 0, ~\langle \xi (t) \xi (t') \rangle = 2 \gamma T_0 \delta (t-t'), \\
\left[1\right] {{\scriptstyle\alpha \atop \displaystyle\rightharpoonup} \atop {\displaystyle\leftharpoondown \atop \scriptstyle\alpha}} \left[2\right].
\end{array}
\label{eq:flashing_ratchet}
\end{equation}
We assume that the mass $m$ and viscosity $\gamma$ of the particle are the same for each state.
$\xi (t)$ is white Gaussian thermal noise with a magnitude given by the temperature of the heat bath $T_0$.

\begin{figure}[tbp]
\begin{center}
\scalebox{1.0}{\includegraphics{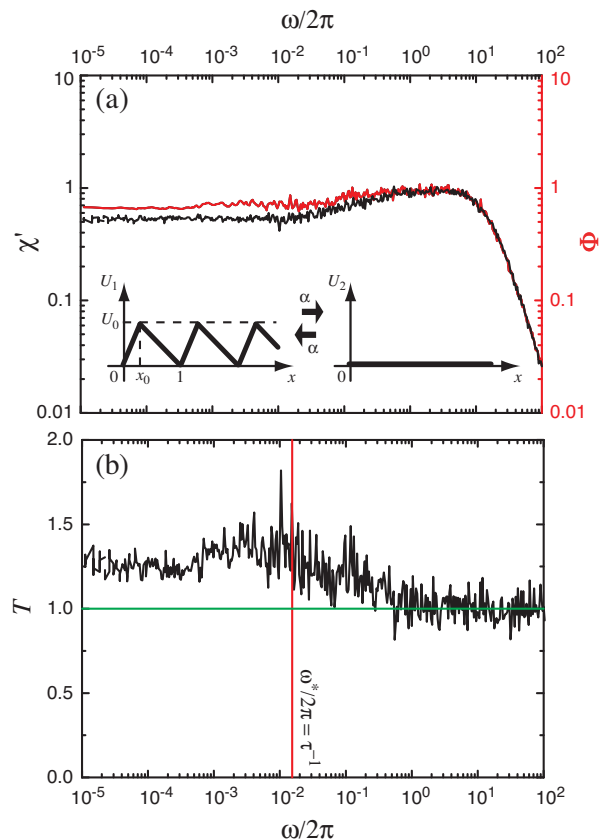}}
\caption{Numerical result for the dissipation-temperature computation for a flashing ratchet model (color). (a) In-phase susceptibility $\chi'[\omega]$ and two-sided power spectrum density $\Phi[\omega]$. The black and red lines indicate the susceptibility and the power spectrum, respectively. Inset : Scheme for a flashing ratchet. The parameters used here were $U_0 = 5$, $x_0 = 0.2$, $\alpha = 0.5$, mass of the particle $m = 0.01$, viscous coefficient $\gamma = 1$, and the heat bath temperature $T_0 = 1$. (b) The dissipation temperature $T[\omega]$ calculated from the data in (a). The horizontal green line represents the temperature of the heat bath $T_0$. The vertical red line is the inverse of the mean passage time $\tau = 65.0\pm0.874$, which was also computed numerically.}
\label{fig1}
\end{center}
\end{figure}

We numerically solved the equations (\ref{eq:flashing_ratchet}) with some initial conditions.
After some time lag, the system reaches a steady state.
First, the double-sided power spectrum density is calculated by Fourier transformation for the velocity $v(t) = \dot{x}(t)$ of the particle under no perturbation, i.e. $f=0$.
By the Wiener-Khinchine theorem \cite{Kubo}, this amount is equal to $\Phi [\omega]$.
Next, the velocity is determined in the presence of a small oscillating force $f(t) = \epsilon \cos (\omega t)$. The average linear power loss $f(t) (v(t) - v_0)$ results in the in-phase susceptibility $\chi' [\omega] = a \langle f(t) (v(t) - v_0) \rangle / \epsilon^2$, where $a = 1$ for $\omega = 0$ and $a = 1/2$ for $\omega \neq 0$ (see Fig. 1 (a)).
Finally, we can obtain the dissipation temperature by Eq. (\ref{eq:FDT_temp}).
We confirmed that the dissipation temperature calculated in this procedure equals $T_0$ in the case of $\alpha = 0$, i.e. in equilibrium.
Figure 1 (b) shows the numerically computed dissipation temperature $T[\omega]$ in a nonequilibrium case.
As shown, the dissipation temperature has different values with a long time limit ($\omega \to 0$) and a short time limit ($\omega \to \infty$), and undergoes a sigmoid-like transition at an intermediate frequency $\omega^*$.
The dissipation temperature equals the temperature of the heat bath in the short time limit and increases with a long time limit \cite{Harada}.
The transition timescale $2 \pi/\omega^*$ coincides the mean passage time $\tau$ of the particle in this ratchet system, i.e. the mean time to pass along a single period  of the potential.

We now address the energetics in terms of this dissipation temperature.
We assume that the motion of a ratchet system or a molecular motor can be approximated by the following generalized Langevin equation:
\begin{equation}
\int_{-\infty}^{t} \Gamma(t-t') v(t') \mathrm{d}t' = - f + \xi(t) + \Xi(t).
\label{eq:Langevin}
\end{equation}
$-f$ is a small external load. $\xi(t)$ is a thermal noise which satisfies normal FDT as $\langle \xi(t) \rangle = 0$ and $\langle \xi(t) \xi(t') \rangle = T_0 \Gamma(t-t')$.
Suppose that $\Xi(t)$ is the nonequilibrium portion of noise with the mean value $\langle \Xi(t) \rangle = \Xi_0$ and the correlation $\langle (\Xi (0) - \Xi_0) (\Xi (t) - \Xi_0) \rangle = M (t-t')$. We assume no correlation between $\xi(t)$ and $\Xi(t)$.

By solving Eq. (\ref{eq:Langevin}) with Fourier transformation and comparing the results with the definition of $\chi[\omega]$, we obtain
\begin{equation}
\chi[\omega]^{-1} = \Gamma[\omega] \equiv \int_{0}^{\infty} \Gamma(t) e^{i \omega t} \mathrm{d}t,
\label{eq:sus_friction}
\end{equation}
where $\Gamma[\omega]$ is a Fourier-Laplace transformation of the friction function $\Gamma(t)$.
Similarly, the following relation is obtained,
\begin{equation}
\Phi[\omega] = (T_0 \Gamma'[\omega] + M[\omega])/|\Gamma[\omega]|^2,
\label{eq:corr_mag}
\end{equation}
where $M[\omega] \equiv 1/2 \int_{-\infty}^{\infty} M(t) e^{i \omega t} \mathrm{d}t$, and $\Gamma'[\omega] \equiv \mathrm{Re} \Gamma[\omega]$.
From Eq. (\ref{eq:FDT_temp}) (note that $\chi'[\omega] = \Gamma'[\omega]/|\Gamma[\omega]|^2$), 
\begin{equation}
T[\omega] = T_0 + \frac{M[\omega]}{\Gamma'[\omega]}.
\label{eq:FDT2}
\end{equation}
This relation corresponds to the 2nd kind of FDT.
We define $\Delta T[\omega] \equiv T[\omega] - T_0 = M[\omega]/\Gamma'[\omega]$.

We now perform long-time averaging on Eq. (\ref{eq:Langevin}) after multiplying by $v(t)$ and obtain the expression
\begin{eqnarray}
\lefteqn{f \langle v(t) \rangle = - \int_{-\infty}^{t} \Gamma(t-t') \langle v(t) v(t') \rangle \mathrm{d}t'} \\ \nonumber
& & {}+ \int_{-\infty}^{\infty} \frac{\Gamma'[\omega]}{\Gamma[\omega]} T_0 \frac{\mathrm{d}\omega}{2 \pi} + \Xi_0 \langle v(t) \rangle + \int_{-\infty}^{\infty} \frac{\Gamma'[\omega]}{\Gamma[\omega]} \Delta T[\omega] \frac{\mathrm{d}\omega}{2 \pi}.
\label{eq:energy_balance}
\end{eqnarray}
This expression represents a balance of energy flux.
The left-hand side represents the work against the external load.
On the right-hand side, the 1st term corresponds to dissipation into the heat bath, the 2nd term is the energy input from the bath, and the 3rd and 4th terms reflect external energy input.
Therefore, the efficiency of energy transduction from external energy to work can be defined as
\begin{equation}
e \equiv \frac{f \langle v \rangle}{Q_{\mathrm{ir}} + \Xi_0 \langle v \rangle},
\label{eq:efficiency}
\end{equation}
where $Q_{\mathrm{ir}} \equiv \int_{-\infty}^{\infty} \left(\Gamma'[\omega]/\Gamma[\omega]\right) \Delta T[\omega] \mathrm{d}\omega/2 \pi$.
It is clear from this expression that a dissipation temperature greater than the heat bath temperature leads to irreversible dissipation and decreases the efficiency.
On the other hand, in the ideal case where $\Delta T [\omega] = 0$ and $f \to \Xi_0$, which corresponds to a quasistatic process, we get $e \to 1$.
Thus, Eq. (\ref{eq:efficiency}) represents how much work can be extracted from thermodynamical free energy \cite{Derenyi}.
This definition of efficiency must be identified from what usually appears in the theory for the Carnot cycle in thermodynamics, $\eta$, which reflects how much work can be extracted from an absolute amount of input energy. In the present context, $\eta$ is expressed as $\eta \equiv f \langle v \rangle/\left( Q + \Xi_0 \langle v \rangle \right)$, where $Q \equiv \int_{-\infty}^{\infty} \left(\Gamma'[\omega]/\Gamma[\omega]\right) T[\omega] \mathrm{d}\omega/2 \pi$.

It is also clear that the condition $Q_{\mathrm{ir}} \geq 0$ is necessary because the efficiency $e$ must be less than unity.
Thus, the fact that the dissipation temperature is equal to or greater than the heat bath temperature is associated with the stability of the nonequilibrium steady state.

Let us compare this result with the experimental results for the conventional kinesin.
Although neither the correlation function nor susceptibility have been measured for kinesin, information is available regarding the diffusion coefficient, mobility and so on.
Thus, we adopt the following approximation based on the assumption that the dissipation temperature of kinesin has the same characteristics as that of a flashing ratchet (Fig. 2 (b)).
First, we suppose that the dissipation temperature of kinesin is also a sigmoidal function with a transition around the mean passage time $\tau$.
We also suppose that the imaginary part of $\Gamma [\omega]$, which is related to the inertia of the motor, is negligible above the timescale of inertia $m/\gamma$, which is much smaller than $\tau$.
Under these assumptions, $Q_{\mathrm{ir}}$ can be approximated as $Q_{\mathrm{ir}} \approx 2 \Delta T[0] \omega^* / 2 \pi = 2 (D/\mu - T_0)/\tau$, where $D$ and $\mu$ are the diffusion coefficient and mobility, respectively.
Thus, efficiency can be expressed as
\begin{equation}
e \approx \frac{\mu f (\Xi_0 - f)}{2(D/\mu - T_0)/\tau + \mu \Xi_0 (\Xi_0 - f)}.
\label{eq:approx_eff}
\end{equation}

We used the values shown in Table 1 from Refs. \cite{Svoboda, Nishiyama}.
\begin{table}[tdp]
\caption{Physical parameters of kinesin under a saturated ATP concentration ($\ge 1$ mM) at room temperature \cite{Svoboda, Nishiyama}. $\mu$, $D$, $\tau$, and $\Xi_0$ represent the mobility, the diffusion coefficient, the mean passage time, and the internal force, respectively. In these experiments, a small polymer particle attached to kinesin molecules was trapped with optical tweezers and the movement of the particle on a microtubule was observed at nano-meter and milli-second resolution  \cite{exp}.}
\begin{center}
\begin{ruledtabular}
\begin{tabular}{cccc}
$\mu$ (nm/s$\cdot$pN)& $D$ ($\mathrm{nm^2}$/s) & $\tau$ (ms) & $\Xi_0$ (pN) \\
\hline
100 & 1300 & 10 & 8 \\
\end{tabular}
\end{ruledtabular}
\end{center}
\label{table}
\end{table}
The value of $\Xi_0$ was determined from a stalling force for kinesin, since the force-velocity curve of kinesin is almost linear.
Based on these values, the long-time dissipation temperature of kinesin is $T[0] = D/k_{\mathrm{B}}\mu \approx 900$ K, while $T_0 \approx 300$ K (We only get back Boltzmann constant $k_{\mathrm{B}}$ in this paragraph).
Equation (\ref{eq:approx_eff}) can be evaluated as $e \approx f (8-f)/\left(16.8 + 8\cdot(8-f)\right)$.
Figure 2 shows this function along with the experimental data for kinesin taken from Ref. \cite{Nishiyama}.
These data agree very well without any fitting parameters.

\begin{figure}[tbp]
\begin{center}
\scalebox{1.0}{\includegraphics{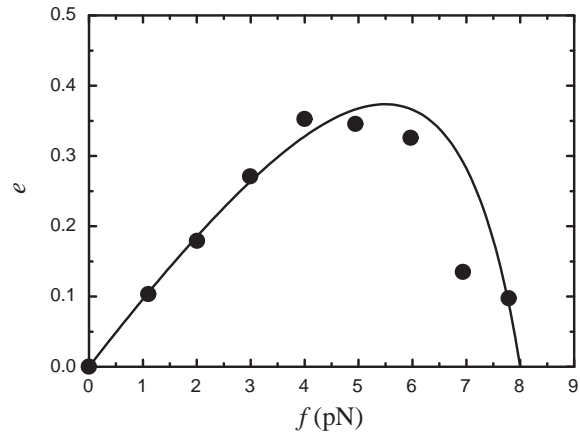}}
\caption{Energetic efficiency against a load force for kinesin. The solid line shows the approximation given by equation (\ref{eq:approx_eff}) with known physical values for kinesin. Black points are the experimental data from Ref. \cite{Nishiyama}.}
\label{fig1}
\end{center}
\end{figure}

A remarkable feature of the present theory is its closed formalism only with observables. We assumed no specific model to calculate the efficiency.
However we need to discuss the assumption used here: the generalized phenomenological Langevin equation (\ref{eq:Langevin}).
First, we must carefully consider whether the motion of a molecular motor is a Markovian process that can be described with such a linear Langevin equation. This assumption requires that the internal force of the motor $\Xi(t)$ must be small enough. For the same reason, the present theory is limited on the case of a small external load $-f$ \cite{nonlinearity}. This assumption seems to be satisfied for kinesin, since its response to a external force is linear.
Moreover, although we assumed an equation only for velocity, it is necessary that there is little or no dissipation for the other degrees of freedom, so that the above assumption does not affect the energetics.
If this point is violated, an argument for another degree of freedom is conducted in parallel with the argument presented here, by measuring the dissipation temperature for such a degree of freedom and constructing a suitable generalized Langevin equation.
In this case, the dissipation temperature will be a tensor (we do not present the details here).
However, the fact that the present theory which considers only velocity can reproduce the experimental data of kinesin might mean that dissipation for the other degrees of freedom is negligible compared to that for velocity.

In conclusion, we have presented an expression for the energetic efficiency of a molecular motor using observable quantities alone. A concept of the dissipation temperature is also introduced based on the susceptibility and the correlation function of the motor. The dissipation temperature for a flashing ratchet exhibits two regimes with respect to the observation timescale. In the short-time regime, the dissipation temperature equals the temperature of the heat bath, while it increases in the long-time regime. The two regimes meet around the mean passage time of the ratchet system.
The expression of efficiency derived in the present paper showed a good agreement with the experimental data for kinesin.

The dissipation temperature is an observable quantity in principle using recently developed single-molecule-detection techniques for molecular motors, although there has been no previous observation.
The measurement of the dissipation temperature of processive motors, such as kinesin, or cooperative systems, such as muscle fibers or flagella, should be an interesting problem in itself, to clarify their energetic properties over a wide range of timescale.
The data will be also useful for testing previously described models of molecular motors.

The issue of effective temperature is also related to fundamental problems of thermodynamics in nonequilibrium systems.
It has been recently reported that the effective temperature based on FDT actually has the nature of temperature in several nonequilibrium systems \cite{Barrat, Sasa}.
We need a further investigation to judge whether the dissipation temperature discussed here can also be a natural temperature for molecular motors, which hopefully leads us to an understandng of nonequilibrium thermodynamics for molecular motors.

The author acknowledge useful discussions with Prof. K. Yoshikawa, Dr. S. Sasa, and Ms. Hayashi.
This work is supported in part by Research Fellowships of the Japan Society for the Promotion of Science for Young Scientists (No. 05494).


\begin{thebibliography}{99}
\bibitem{Block} K. Svoboda and S. M. Block, Cell \textbf{77}, 773 (1994).
\bibitem{Reimann} P. Reimann, Phys. Rep. \textbf{361}, 57 (2002).
\bibitem{Parrondo} J. M. R. Parrondo and B. J. de Cisneros, Appl. Phys. A \textbf{75}, 179 (2002).
\bibitem{Kubo} R. Kubo, M. Toda, and N. Hashitsume, \textit{Statistical Physics II: Nonequilibrium Statistical Mechanics} (Springer, Berlin, 1991).
\bibitem{Peliti} L. F. Cugliandolo, J. Kurchan, and L. Peliti, Phys. Rev. E \textbf{55}, 3898 (1997).
\bibitem{Astumian} R. D. Astumian and M. Bier, Phys. Rev. Lett. \textbf{72}, 1766 (1994).
\bibitem{Prost} J. Prost, J. -F. Chauwin, L. Peliti, and A. Ajdari, Phys. Rev. Lett. \textbf{72}, 2652 (1994).
\bibitem{Okada} Y. Okada and N. Hirokawa, Science \textbf{283}, 1152 (1999).
\bibitem{Harada} The long-time dissipation temperature is equal to the ratio of the diffusion coefficient to the mobility of a particle, which has recently been investigated in an experimentally realized thermal ratchet. [T. Harada and K. Yoshikawa, \textit{submitted}].
\bibitem{Derenyi} I. Der\'{e}nyi, M. Bier, and R. D. Astumian, Phys. Rev. Lett. \textbf{83}, 903 (1999).
\bibitem{Svoboda} K. Svoboda, P. P. Mitra, and S. M. Block, Proc. Natl. Acad. Sci. USA \textbf{91}, 11782 (1994).
\bibitem{Nishiyama} M. Nishiyama, H. Higuchi, and T. Yanagida, Nat. Cell. Biol. \textbf{4}, 790 (2002).
\bibitem{nonlinearity} On the contrary, bottom-up type theories \cite{Parrondo} remains valid in a nonlinear regime [see, e.g.: K. Sekimoto, J. Phys. Soc. Jpn. \textbf{66}, 1234 (1997)].
\bibitem{Barrat} L. Berthier and J.-L. Barrat, Phys. Rev. Lett. \textbf{89}, 095702 (2002).
\bibitem{Sasa} K. Hayashi and S. Sasa, cond-mat/0309618.
\bibitem{exp} There have been many reports on the measurement of diffusion coefficients by single-molecule imaging techniques. [See, e.g.: Y. Inoue, A. H. Iwane, T. Miyai, E. Muto, and T. Yanagida, Biophys. J. \textbf{81}, 2838 (2001)]. In these measurements, the lower limit for the value of the diffusion coefficient should be on the order of several thousands $\mathrm{nm^2/s}$, since the spatial resolution of the optical method is at best about 100 nm and  a motor molecule runs for about 2 - 3 s. However, this limitation may be too large, since the value measured by the nanometry technique \cite{Svoboda} was much smaller. Therefore, we used the data obtained by nanometry.
\end{thebibliography}
\end{document}